\documentclass{llncs}

\usepackage[utf8x]{inputenc}
\usepackage{color}
\usepackage[sf,SF]{subfigure}
\usepackage{graphicx}

\begin{document}

\title{Dependability in Aggregation by Averaging}

\bibliographystyle{unsrt}

\author{Paulo Jesus \and Carlos Baquero \and Paulo Sérgio Almeida}
\institute{University of Minho (CCTC-DI)\\
Campus de Gualtar, 4710-057 Braga, Portugal\\
\email{\{pcoj, cbm, psa\}@di.uminho.pt}}

\maketitle

\begin{abstract}

Aggregation is an important building block of modern distributed applications, allowing the determination of meaningful properties (e.g. network size, total storage capacity, average load, majorities, etc.) that are used to direct the execution of the system. In the recent years, several approaches have been proposed to compute aggregation functions on distributed settings, exhibiting different characteristics, in terms of accuracy, time and communication complexity. However, the majority of the existing aggregation algorithms exhibit relevant dependability issues, when prospecting their use in real application environments. 
In this paper, we reveal some dependability issues of aggregation algorithms based on iterative averaging techniques, giving some directions to solve them. This class of algorithms is considered robust (when compared to common tree-based approaches), being independent from the used routing topology and providing an aggregation result at all nodes. However, their robustness is strongly challenged and their correctness often compromised, when changing the assumptions of their working environment to more realistic ones. The correctness of this class of algorithms relies on the maintenance of a fundamental invariant, commonly designated as \emph{mass conservation}. We will argue that this main invariant is often broken in practical settings, and that additional mechanisms and modifications are required to maintain it, incurring in some degradation of the algorithms performance.
In particular, we discuss the behavior of three representative algorithms (\emph{Push-Sum Protocol}~\cite{Kempe:2003p700}, \emph{Push-Pull Gossip protocol}~\cite{Jelasity:2005p664} and \emph{Distributed Random Grouping}~\cite{JenYeuChen:2006p3916}) under asynchronous and faulty (with message loss and node crashes) environments.  More specifically, we propose and evaluate two new versions of the \emph{Push-Pull Gossip protocol}, which solve its message interleaving problem (evidenced even in a synchronous operation mode).  

\end{abstract}

\section{Introduction}
\label{sec:intro}

With the advent of multi-hop ad-hoc networks, sensor networks and large
scale overlay networks, there is a demand for tools that can abstract
meaningful system properties from given assemblies of nodes. In such
settings, aggregation plays an essential role in the design of distributed
applications~\cite{Renesse:2003p677}, allowing the determination of
network-wide properties like network size, total storage capacity, average
load, and majorities. Although apparently simple, in practice aggregation
has reveled itself to be a non trivial problem, especially when seeking
solutions in distributed settings, where no single element holds a global
view of the whole system.

In the recent years, several algorithms have addressed the problem from
diverse approaches, exhibiting different characteristics in terms of
accuracy, time and communication tradeoffs. A useful class of aggregation
algorithms is based on \emph{averaging} techniques. Such algorithms start from a
set of input values spread across the network nodes, and iteratively average
their values with active neighbors. Eventually all nodes will converge to
the same value and can estimate some useful metric. Averaging 
techniques allow the derivation of different aggregation functions besides
average (like count, sum, maximum and minimum), according to the initial
combinations of input values. E.g., if one node starts with input $1$ and
all other nodes with input $0$, eventually all nodes will end up with the
same average $1/n$ and the network size $n$ can be directly estimated by all
of them~\cite{Jelasity:2004p662}.

The main objective of this work is to expose relevant dependability issues
of existing aggregation by averaging algorithms, when challenged by
practical implementations in realistic scenarios. For this purpose, we
discuss and evaluate the behavior of three representative averaging
algorithms, when confronted with practical concerns like communication
asynchrony, message loss and node failure. We choose to analyze the
following algorithms: \emph{Push-Sum Protocol}~\cite{Kempe:2003p700} (PSP),
\emph{Push-Pull Gossip protocol}~\cite{Jelasity:2005p664} (PPG), and
\emph{Distributed Random Grouping}~\cite{JenYeuChen:2006p3916} (DRG). To the
best of our knowledge this is the first evaluation of averaging algorithms
focusing on dependability and taking into account practical implementation
concerns.

The remaining of this paper is organized as follows. We briefly refer to the
related work on aggregation algorithms in Section~\ref{sec:related_work}. A
detailed analysis of some representative averaging aggregation algorithms,
concerning their practical implementation on real distributed systems, is
discussed in Section~\ref{sec:analysis}. In Section~\ref{sec:case_study}, we
propose two solutions to fix the interleaving issues exhibited by PPG, and
compare them with the original algorithm in a common simulation environment.
Finally, we make some concluding remarks in Section~\ref{sec:conclusions}.

\section{Related Work}
\label{sec:related_work}

Several aggregation algorithms have been proposed in the last years,
tackling the problem for different settings, and yielding different
characteristics in terms of accuracy, time and communication complexity.

Classical approaches, like TAG~\cite{Madden:2002p2402}, perform a tree-based
aggregation where partial aggregates are successively computed from child
nodes to their parents until the root of the aggregation tree is reached
(requiring the existence of a specific routing topology). This kind of
aggregation technique is often applied in practice to Wireless Sensor
Network (WSN)~\cite{Madden:2002p695}. Other tree-based aggregation
approaches can be found in \cite{Li:2005p685}, and \cite{Birk:2006p3907}. We
should point out that, although being energy-efficient, the reliability of
these approaches may be strongly affected by the inherent presence of
single-points of failure in the aggregation structure. 

Another common class of distributed aggregation algorithms is based on
averaging
techniques~\cite{Kempe:2003p700,Jelasity:2005p664,Jelasity:2004p662,JenYeuChen:2006p3916,Wuhib:2007p832};
Here, the values of a variable across all nodes are averaged iteratively.
This kind of approaches is independent from the routing topology, often
using a gossip-based communication scheme between peers. In this study, we
will specifically discuss three of these approaches:
PSP~\cite{Kempe:2003p700}, PPG~\cite{Jelasity:2005p664}, and
DRG~\cite{JenYeuChen:2006p3916}.

Alternative aggregation algorithms based on the application of
\emph{probabilistic} methods, can also be found in the literature. This is
the case of Extrema Propagation~\cite{BAM09} and
COMP~\cite{MoskAoyama:2006p686}, which reduce the computation of an
aggregation function to the determination of the minimum/maximum of a
collection of random numbers. These two techniques tend to emphasize speed,
being less accurate than averaging approaches.

Specialized probabilistic algorithms can also be used to compute specific
aggregation functions, such as \textsc{count} (e.g. to determine the network
size). This type of algorithms essentially relies on the results from a
sampling process to produce an estimation of the aggregate, using properties
of random walks, capture-recapture methods and other statistic
tools~\cite{Massoulie:2006p4521,Ganesh:2007p745,Mane:2005p659,Kostoulas:2005p682}.

\section{Analysis}
\label{sec:analysis}

We analyze the practical implementation of aggregation algorithms based on
averaging techniques, envisioning their deployment on real distributed
network systems (e.g WSN and P2P overlay networks). In particular, we
discuss three representative algorithms from this class:
PSP~\cite{Kempe:2003p700}, PPG~\cite{Jelasity:2005p664}, and
DRG~\cite{JenYeuChen:2006p3916}. The performed analysis focuses on the
reliability of these algorithms, in order to provide an accurate aggregation
estimate, on realistic application scenarios, which are commonly governed by
communication asynchrony, and failures. More specifically, this analysis is
confined to four main practical settings/concerns: 

\begin{enumerate} 

\item \textbf{Synchronous model} -- refers to the common synchronous operation mode,
where the algorithms execution proceeds in lock-step rounds and without
faults. In practice, where networks are typically asynchronous, it is
possible to implement synchrony over an asynchronous fault-free network (see
Chapter 16 of \cite{Lynch}), using a \emph{synchronizer}. Notice that even
under this strong synchrony assumption, algorithms that span more than one
round, may see their messages interleaved across rounds; 

\item \textbf{Asynchronous model} -- in these settings, message transit takes a
finite but unknown time. We consider that nodes communicate using FIFO
channels, and no faults will occur. This means that no message in transit
can be surpassed by any other message from the same source. 
Commonly, in practice, the transport communication layer (e.g. TCP) can
provide a reliable and ordered message delivery, retransmitting undelivered
data packets and using a sequence number to identify their order. Like in
previous settings, interleavings may also frequently occur; 

\item \textbf{Message loss} -- corresponds to the loss of communication
data, due to a temporary link or node failure. In this settings, we consider
asynchronous non-FIFO communication, where no guarantee on message
delivery is made (like in UDP); 

\item \textbf{Node Crash} -- refers to the permanent failure of a node at an
arbitrary time  -- \emph{crash-stop} model. If a node crashes, it will no
longer receive nor send messages, and will be considered as permanently
unavailable from that time on. 

\end{enumerate}

	Aggregation algorithms based on averaging are independent from the
network routing topology, and able to produce an estimate of the resulting
aggregate at every network node. The main principle of this kind of
algorithms is based on an iterative averaging process between small 
sets of nodes. Eventually, all nodes will converge to the correct value by
performing the averaging process among all the network. 
	
	The correctness of averaging aggregation algorithms depends on the
maintenance of a fundamental invariant, commonly designated as the ``mass
conservation''. This property states that the sum of the aggregated values
of all network nodes must remain constant along the algorithm's execution,
in order for it to converge to the correct result~\cite{Kempe:2003p700}.
	
	This kind of aggregation technique intends to be more robust than
classical tree-based approaches, by removing the dependency from a specific
routing topology, and providing an estimation of the aggregate at every
node. For instance, these algorithms are often based on a gossip (or
epidemic) communication scheme, which is commonly thought to be robust.
Although, similarly to gossip communication
protocols~\cite{Alvisi:2007p3354}, the robustness of aggregation algorithms
can be challenged, according to the assumptions made on the environment in
which they operate. 

	Next, we discuss and expose dependability issues of some aggregation
algorithms operating on a fixed network, but under more realistic
assumptions, such as: asynchronous message exchanges, link and node
failures. 
	
\subsection{Push-Sum Protocol}
\label{sec:push-sum}

The PSP~\cite{Kempe:2003p700} is a simple gossip-based aggregation
algorithm, essentially consisting on the distribution of shares across the
network. Each node maintains and iteratively propagates information of a
\emph{sum} and a \emph{weight}, which are sent to randomly selected nodes.
In more detail: at each time step $t$ (synchronous round), each node $i$
sends to a target node (chosen uniformly at random) and to itself, a pair
$(\frac{1}{2}s_{t,i}, \frac{1}{2}w_{t,i})$ containing half of its current
sum $s_{t,i}$ and weight $w_{t,i}$; the values $s_{t,i}$ and $w_{t,i}$ are
updated by the sum of all contributions received in the previous time step
($t-1$); an estimate of the aggregation result can be provided at all nodes
by the ratio $s_{t,i}/w_{t,i}$. 

Distinct aggregation functions can be computed with these scheme, by varying
only on the starting values of the \emph{sum} and \emph{weight} variables at
all nodes. For instance, considering an initial input value $x_i$ at each
node $i$, the following functions can be computed, resorting to distinct
initializations: \textsc{average} ($s_{0,i} = x_i$ and $w_{0,i} = 1$ for all
nodes); \textsc{sum}  ($s_{0,i} = x_i$ for all nodes, only one node sets
$w_{0,i} = 1$ and the remaining $w_{0,i} = 0$); \textsc{count} ($s_{0,i} =
1$ for all nodes, only one with $w_{0,i} = 1$ and the others $w_{0,i} = 0$).

The continuous execution of this protocol throughout all the network allows
the eventual convergence of all nodes estimates to the correct aggregation
value, as long as none of the exchanged values are lost. As stated by the
authors, the correctness of the algorithm relies on the \emph{mass
conservation} property. In particular, when no messages are in transit, for
any time step $t$, the value $\sum_{\forall i} \frac{s_{t,i}}{w_{t,i}}$ is
constant. 

Aware from the crucial importance of this invariant, Kempe et al.
considered a variation of the algorithm to cope with message loss and the
initial failure of some nodes\footnote{\emph{Initial failure}, refers to
nodes that have failed from the beginning of the computation.}. They assume
that all nodes possess the ability to detect when their messages did not
reach their destination, and modified the algorithm to send the undelivered
data to the node itself, in order to recover the lost ``mass''. Furthermore,
they assume that a node can only orderly leave the network, after sending
all sums and weights to its peers, not supporting node crashes. In light of
these stated assumptions, we extend the discussion of PSP under more
realistic settings. 

\subsubsection{Synchronous model}
The proposed algorithm guarantees the maintenance
of the mass conservation invariant on a synchronous execution model without
faults, assuring its correctness -- convergence to the true aggregation
result. 

\subsubsection{Asynchronous model}

No issue was identified under asynchronous settings. Message delays may
reduce the convergence speed of the algorithm, but will not compromise its
correctness (as long as no message is lost). If at some arbitrary point $t$
we stop the execution of the algorithm and wait for all message to be
received and processed, we can verify that the ratio $s_{t,i}/w_{t,i}$ will
meet the mass conservation property (the value $\sum_{\forall i}
\frac{s_{t,i}}{w_{t,i}}$ will be equal to the initial value $\sum_{\forall
i} \frac{s_{0,i}}{w_{0,i}}$).

\subsubsection{Message Loss}

As referred by the authors, in order to support message loss, independently
from its cause (temporary link/node failure, or initial permanent failures
that makes some nodes unreachable), they assume that each node is able to
detect messages that have not reached their destination. In this case, the
lost mass will be re-sent to the source node itself. This is a very
unrealistic assumption. Using an acknowledgement-based scheme to infer message loss, as suggested, would amounts to solving the \emph{coordinated attack problem}, which in an asynchronous model under possible message loss has been shown to be impossible \cite{Gray:1978}. Furthermore, even if it were
possible, it would introduce additional waiting delays in the protocol, in
order to receive a delivery notification for each sent message.

\subsubsection{Node Crash}

This algorithm does not support node crash. In order to maintain the
correctness of the aggregation process, nodes will have to leave the network
neatly, after sending all their mass to another node. Such optimistic
assumption cannot be often used in practice, since node failures are not likely 
to be predicted. Nevertheless, one could consider a mechanism in which all
the nodes state will be consistently replicated (at each neighbor), enabling
alive nodes to subsequently recover the ``mass'' from a crashed node. 

In order to be robust against node failures, G-GAP~\cite{Wuhib:2007p832}
extended the PSP, implementing a scheme based on the computation of recovery
shares and the explicit acknowledgement of mass exchanges between peers.
However, this approach provides only a partial support against this type of
faults, supporting discontiguous node crashes (assuming that two adjacent
nodes do not fail within a short time period of each other).

\subsection{Push-Pull Gossip}
\label{sec:push-pull}

The PPG protocol described in \cite{Jelasity:2004p662} and
\cite{Jelasity:2005p664} is based on an anti-entropy aggregation process,
being quite similar to PSP. The main difference of this algorithm relies on
its \emph{push-pull} process, which enforces a symmetric pairwise mass
exchange between peers. The induction of this action/reaction pattern,
coerces the average settlement between pairs, intending to immediately
resolve the differences between nodes. The iterative execution of this
push-pull process across all the network will provide the convergence of the
algorithm to the correct value (faster, when compared to PSP). 

Periodically, each node sends its current aggregation value to another node
chosen uniformly at random -- \emph{push}, and waits for the response with
the value from the target node -- \emph{pull} -- to further apply the
aggregation function, obtaining a new estimate of the aggregate. Each time a
node receives a push message, it sends back its current value, and only then
computes the new aggregate, averaging the received and sent value.  At the
end of a push-pull operation, both involved nodes will yield the same
result. 

In \cite{Jelasity:2004p662} the authors do not address mass conservation
issues (like the impact of link/node failures), focusing on the efficient
implementation of a pair selection method, and considering the extension of
the algorithm with a restarting mechanism (executing the protocol during a
predefined number of cycles, depending on the desired accuracy) in order to
be adaptive and handle network changes (nodes joining/leaving). The
subsequent study in~\cite{Jelasity:2005p664} proposed a solution for the
gossip-based aggregation directed to its practical applicability,
considering some modifications to cover some practical issues, like: using
timeouts to detect possible faults, ignoring the data exchanges in those
situations; executing several instances of the algorithm in parallel to
reduce the effect of message loss; making use of a restart mechanism to cope
with node crashes.

We should point out that along the discussion carried out by the authors,
they intrinsically assume that the core of the push-pull process is atomic,
also referred as ``the variance reduction step'' ($w_i=w_j=(w_i+w_j)/2$). In
practice, additional modifications must be considered to guarantee the
atomicity of \emph{push-pull}. Like common aggregation algorithms by
averaging, the correctness of PPG depends from the maintenance of the mass
conservation property. Due to its core resemblance with PSP, one could expect
to find similar practical issues when prospecting its use in real
application scenarios. However, the push-pull process will introduce
additional atomicity constraints, that will restrain the use of this
algorithm even under a synchronous operation mode. 

\begin{figure}
\centering
\includegraphics[width=0.42\textwidth]{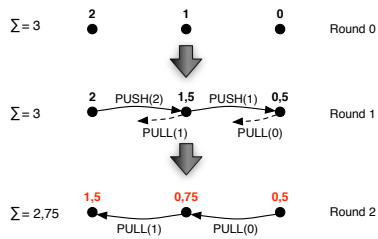}
\caption{Violation of the mass conservation invariant in the Push-Pull Gossip protocol, due to message interleaving (approximation values at the end of each round).} \label{fig:ppg_issue}
\end{figure}

\subsubsection{Synchronous model}

Even in synchronous settings (lock-step execution), the message delivery order
may affect the convergence to the true result of PPG. It is fundamental to
guarantee the atomicity of the push-pull process in those settings, in order
to conserve the mass of the system. If a node starts a push-pull and
receives a push message from a third-party node in the middle of the process
(after sending the push and before receiving the pull message), then it will
first update its approximation value according to the third-party data,
before receiving the pull message, and use the result to update its value
again with the pull data. This occurrence will introduce an asymmetry in the
results produced from the push-pull (different values may be computed by
each element), which will incur in a violation of the mass conservation,
depending on the involved values (greater as greater is the discrepancy
between them). Thus, the correctness of the algorithm is affected by message
interleaving, as depicted in Figure~\ref{fig:ppg_issue}. We will detail this
issue in Section~\ref{sec:case_study}, proposing some modifications to the
algorithm to guarantee the push-pull atomicity.

\subsubsection{Asynchronous model}

The interleaving problem of PPG will also be found in asynchronous settings.
Moreover, message delays will increase the possibility to interfere in an
ongoing push-pull operation, and consequently break the mass invariant as
previously described.

\subsubsection{Message Loss}

In practice, considering the independent failure of a push and pull message,
the loss of each one of those messages will yield different effects on the
PPG protocol: if the push message is lost, then no node will update its
state (the source will timeout and the target will not receive the message)
and the mass of the system will be conserved, only slowing down the
algorithm convergence; On the other hand, if the pull message is lost
(meaning that the push message was successfully received), then only the
target node of the exchange process will update its value and the source
will timeout waiting for a response, creating an undesirable asymmetry in
the process and a consequent violation of the mass conservation invariant. 

The authors do not make any assumption regarding the detection of message
loss, neither concerning the restoration of the system mass. Instead of
that, they consider the concurrent execution of multiple instances of the
algorithm, discarding the lower and higher estimates and reporting the
average of the remaining executions as the result. This solution does not
solve the mass conservation issue due to message loss, and only reduces its
impact in the quality of the produced estimate. Further considerations must
be made to ensure the algorithm's correctness in this settings.

\subsubsection{Node Crash}

The removal of nodes from the system originates a violation of the mass
conservation property. In this case, the authors did not show any concern
about recovering the mass changes provoked by unpredictable node crashes
(not distinguishing crashes from nodes voluntarily leaving the network), but
recognize the derivation of a subsequent estimation error. In particular,
they specifically characterize the error of the average estimation as a
function of the constant proportion of nodes failing (at each cycle),
considering that nodes crash before each cycle, and intrinsically assuming
(once more) that each variance reduction step is atomic (which is
unreasonable in practice). 

Indubitably, the introduced estimation error is provoked by the loss of the
values held by crashed nodes, and leads to the convergence of the algorithm
to an incorrect aggregation result. The authors consider the periodic
restart of the PPG algorithm to cope with this issue, reinitializing the
execution of the algorithm with clean input values (restoring the correct
mass of the system). To detect the possible failure of nodes, they consider
the use of a timeout, skipping the exchange step when the timeout expires.
Hopefully, the restart mechanism will minimize the effect of the introduced
convergence error, being transient to each epoch. Nevertheless, this issue
invalidates the continuous use of the algorithm (without restarting), and
may incur in unpredictable approximation results, since the correctness of
the algorithm will be broken.

\subsection{Distributed Random Grouping}
\label{sec:DRG}

A distinct approach based on a Distributed Random Grouping (DRG) is proposed
in \cite{JenYeuChen:2006p3916}. Unlike previous gossip-based aggregation
algorithms, DRG was designed to take advantage of the broadcast nature of
wireless transmissions (where all nodes within radio range will be prone to
hear a transmission), directing its use for WSN. In a nutshell, the
algorithm essentially consists on the continuous creation of random groups
across the network, to successively perform in-group aggregations. Over
time, ensuring that the created groups overlap, the estimated values at all
nodes will eventually converge to the correct network-wide aggregation
result.

DRG defines three different working modes that coordinate its execution:
\emph{leader}, \emph{member}, and \emph{idle}. The algorithm operates as
follows: according to a predefined probability, each node in idle mode can
independently decide to become leader, and broadcast a Group Call Message
(GCM), subsequently waiting for joining members; all (remaining) idle nodes
respond only to the first received GCM with a Joining Acknowledgment (JACK)
tagged with their aggregate estimation, changing their mode to become
members of that group; after gathering the estimate values of group members
from all received JACKs, the leader computes the new group aggregate and
broadcasts the result in a Group Assignment Message (GAM), returning to idle
mode and setting its own estimate with the newly calculated value; each
member of a group waits for the GAM from its leader to update its local
estimate with the result within, not responding to any other request until
then, and returning to idle mode afterwards. 

The performance of this algorithm is highly influenced by its capacity to
create overlapping aggregation groups (size and quantity of groups), which
is defined by the probability of a node to become leader. In terms of
practical concerns, the authors make some considerations about termination
detection for the algorithm, and consider the occurrence of message
collisions and link failures, but only analyze their effect at the initial
stage of the group creation (at the GCM level). In particular, they assume
that link failures only happen between grouping time slots, which is an
unrealistic assumption. A thorough analysis of these issues should be
performed across the algorithm, considering also their impact at the JACK
and GAM level.

\subsubsection{Synchronous model}
DRG will work in synchronous settings, as long as each node is still only
able to join at most one group at each time, which guarantees the
maintenance of the mass conservation property.

\subsubsection{Asynchronous model}

In asynchronous settings, beside the exclusive group entrance, the algorithm
must ensure that each node properly completes its participation in the
group, receiving an aggregation result that has taken into account its
estimated value. Recall that in this model messages are reliably transmitted
in FIFO order but can suffer arbitrary finite delays. 

Since leaders initiating a GCM call cannot anticipate how many nodes will
acknowledge them, there is still a need to consider some timeout for the
reception of the JACK responses.  This opens the possibility that some nodes
that have sent a JACK message will not have this message, and mass,
processed in this iteration of the protocol. It is thus necessary that GAM
messages are augmented with node and iteration ids, so that only nodes whose
JACK reached the leader will consider the subsequent GAM. Iteration ids will
also be important so that leaders discard JACK messages from previous GCMs
that they initiated.  

One can conclude that although there seems to exist no important obstacles
for an implementations under this model, several modifications should be
considered and the correctness of the resulting approach carefully examined.

\subsubsection{Message Loss}

In terms of message loss, the authors consider the occurrence of collisions
and link failures, but they only consider its effect on GMCs, with impacts
on the creation of groups, reducing its expected size. For instance, they
assume that link failures only happen between iterations of the protocol,
which is unrealistic since links can unpredictably fail at any point of the
algorithm execution, provoking the loss of any type of message. 

In this setting, timeouts are needed in the receptions steps for expected
JACK and GAM messages.  The loss of a GCM will have no impact on the
correctness of the algorithm, only preventing nodes from joining the group.
However, loosing a JACK from a node but delivering the subsequent GAM to
that same node will most probably violate mass conservation. The same
happens if GAMs are lost. 

Extensions that assign iteration ids and node ids to the protocol messages
can possibly address some of these issues, but the need to have a
successful GAM reception on nodes that contributed their mass in a
successfully delivered JACK message points again to the
\emph{coordinated attack problem}. 

\subsubsection{Node Crash}

Similarly to previous approaches, this algorithm does not support node
crashes (not addressed by the authors). The impact of the removal of a node
from the network will be proportional to the contribution of its value to
the computation of the global aggregate. As already mentioned before, the
unpredicted failure of a node will alter the total mass of the system,
breaking the mass conservation invariant, and leading the system to converge
to a wrong result.

\section{Case Study: Push-Pull Gossiping}
\label{sec:case_study}

\begin{figure}
\centering
\includegraphics[width=0.5\textwidth]{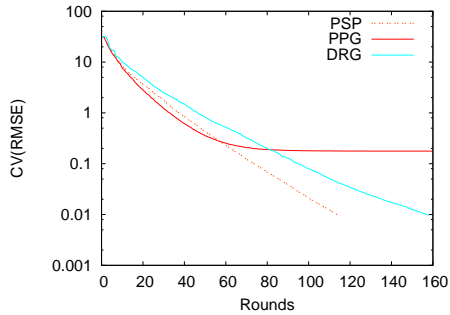}
\caption{Convergence speed of the analyzed algorithms (PSP, PPG and DRG).} \label{fig:all_iter}
\end{figure}

In the previous section, we argue that most aggregation algorithms by
averaging have practical implementation issues in realistic environments. In
order to solve the exposed issues and ensure the algorithms dependability,
important modifications must be introduced to their implementation. However,
these additional modifications will incur in some degradation of their
performance. We now choose to address in more detail PPG, since it is the
only one that does not converge to the true aggregation result in a
synchronous operation mode, as depicted by simulation results of Figure
\ref{fig:all_iter}. 

We use a custom high level simulator to evaluate the execution of the
analyzed algorithms in a synchronous operation mode\footnote{synchronous
model as in Chapter 2 of \cite{Lynch}}, providing a fair
comparison of them. However, the simulation level was detailed enough to
observe some effects on messages passing (in particular, interleaving),
since we assume that all the messages received in a common round will be
processed by the target node in some arbitrary order. 
The depicted results correspond to the average values (number of rounds or messages) required to reach a specific accuracy (coefficient of variation of the root mean square error), obtained from 50
trials of the execution of the algorithms under identical settings. In each
trial different networks with a random topology (generated according to the
Erdős–Rényi model~\cite{Erdos:1960p6048}) and the same characteristics (network of size
1000 and average connection degree of 5) are used. We implement the \textsc{Count}
version of each simulated algorithm (to determine the network size) and tune
all its parameters (e.g. the probability to become leader in DRG) to obtain
its best performance in each simulation scenario.

As previously referred (in Section \ref{sec:push-pull}) and confirmed by
simulations (see Figure \ref{fig:all_iter}), the correctness of PPG (mass
conservation) is affected by the occurrence of messages interleaving in
push-pull operations, occurring even in synchronous rounds settings. In
practice, in order to guaranty the convergence of PPG to the correct result,
an additional property must be considered: the push-pull process must be
atomic, and the approximation value cannot be update by any concurrent
process before the end of an already initiated push-pull procedure. We
propose two concrete alternative versions of the algorithm that tackle this
problem, and guarantee the consistency of the push-pull process:
\emph{Push-Pull Back Cancellation} (PPBC), and \emph{Push-Pull Ordered Wait}
(PPOW).

\begin{figure}
\subfigure[Convergence speed]{\label{fig:pp_iter}\includegraphics[width=0.5\textwidth]{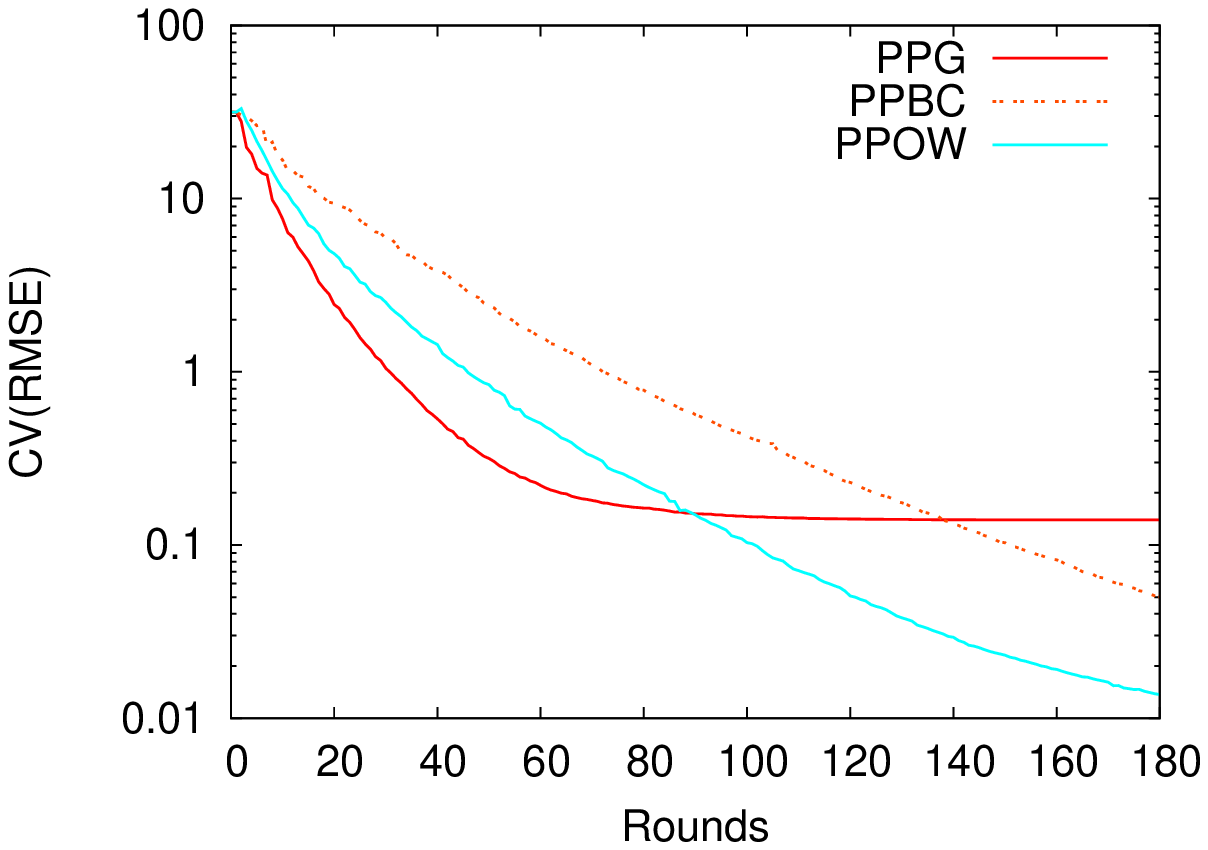}}
\subfigure[Overhead cost]{\label{fig:pp_msg}\includegraphics[width=0.5\textwidth]{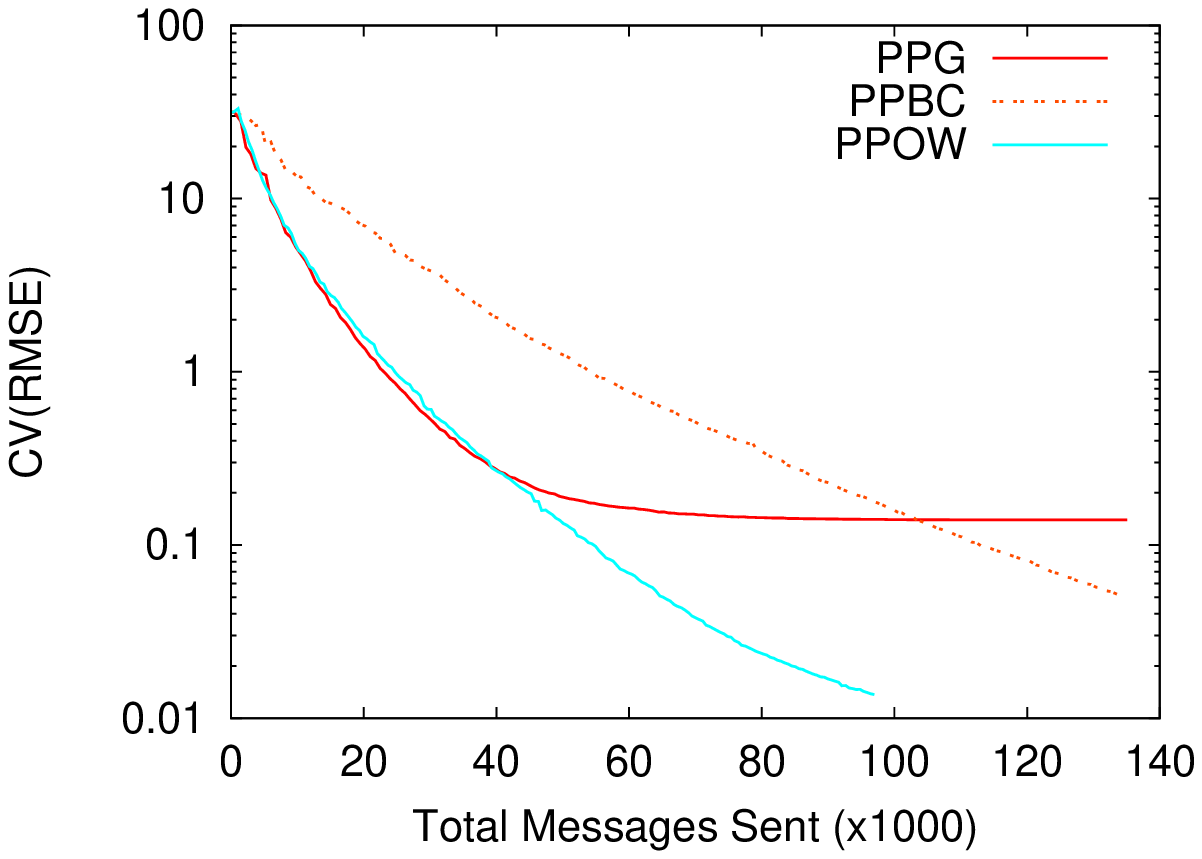}}
\caption{New proposed version of PPG compared to the original algorithm.}
\label{fig:pp}
\end{figure}

\subsection{Push-Pull Back Cancellation}
\label{sec:ppbc}
This version of the algorithm is based on a message cancellation mechanism, to guaranty that no other process will interfere in an ongoing push-pull operation. In this scheme, after sending a \emph{push} message to the chosen target, all the received messages will be ignored (without any local state update) by the source node, until the awaited \emph{pull} message is received. The cancelation mechanism is implemented by sending a \emph{pull} message with the same value received in the \emph{push} one, skipping the correspondent state update. By reflecting back the received approximation value, the update performed by the initiator will not change its approximation value. The main drawback of this mechanism is the execution of dummy push-pulls (that will not contribute to the system convergence), all in order to ensure a consistent state update of ongoing push-pull operations.

\subsection{Push-Pull Ordered Wait}
\label{sec:ppow}

This version of the algorithm adds a message buffering process to the
original version of the algorithm. All \emph{push} messages received while a
push-pull operation has already been initiated by a node are locally
buffered, and will only be processed after the conclusion of the ongoing
push-pull. Notice that, simply considering this buffering mechanism could
incur in deadlocks across the network, due to possible creation of waiting
cycles between nodes. E.g. If node $n_1$ is waiting for node $n_2$, which is
waiting for node $n_3$, and by is turn $n_3$ is waiting for $n_1$, ending
all locked waiting for each other. 

Additional requirements must be considered to avoid this cyclic waiting
situation. For instance, we define a total order between nodes (by setting a
UID to each node), and stipulate that each node can only initiate a
push-pull operation with nodes from a specific order (inferior or superior).
For example, considering the order between nodes $n_1 < n_2 < n_3$ (defined
by their UID, from lower to higher ID values), and imposing that \emph{push}
messages can only be sent to nodes with a higher ID: node $n_3$ could never
be waiting for $n_1$ (breaking the previous cycle), since it has a lower ID
and consequently no \emph{push} messages can be sent to it by $n_3$.
Although, this scheme restrains the execution of push-pull in some
directions, unlike the previous version it does not wast concurrent
push-pulls, and makes the most of their contributions to the convergence of
the algorithm.

\subsection{Simulation Results}
\label{sec:results}

As depicted by Figure \ref{fig:pp}, the modifications introduced to the
original algorithm solve the convergence problem of the algorithm, but at
the cost of a performance degradation in terms of convergence speed (see Figure \ref{fig:pp_iter}). As
expected PPOW performs better than PPBC, since its buffering mechanism
allows the integration of all the received contributions to the aggregation
process. By making all contributions (messages) count, PPOW exhibits similar
results than the original algorithms in terms of overhead (number of message needed to reach a common accuracy), as showed in
Figure \ref{fig:pp_msg}.

\section{Conclusion}
\label{sec:conclusions}

In this study, we expose important implementation and dependability issues
in averaging based aggregation algorithms. Issues that are often overlooked
in the abstract modeling of the algorithms and that must be addressed in any
concrete real scenario.  

In particular, we discuss three established algorithms from this class under
practical settings, namely the \emph{Push-Sum
Protocol}~\cite{Kempe:2003p700}, \emph{Push-Pull Gossip
protocol}~\cite{Jelasity:2005p664} and \emph{Distributed Random
Grouping}~\cite{JenYeuChen:2006p3916}. All algorithms evidence some
dependability issues, that compromise their correctness, when exposed to
asynchronous and faulty (e.g. with message loss or node crash) environments.

Supplementary mechanisms and modifications must be added to the discussed
aggregation algorithms, in order to provide fault tolerance and enable their
practical use. These additional provisions will result in a degradation
of the algorithms performance. In particular, we extend the \emph{Push-Pull
Gossip protocol}, which exhibit issues even on synchronous settings, and
propose two concrete version of the protocol to solve its interleaving
problems: \emph{Push-Pull Back Cancellation}, and \emph{Push-Pull Ordered
Wait}. As showed by the results obtained from simulations, the proposed
algorithms solve the convergence problem of the original algorithm, but
exhibit a worse global performance (especially in terms of convergence
speed).

The depicted vulnerability of current averaging algorithms, and their
importance when seeking high precision aggregates, also served as a
motivation for our ongoing research on \emph{Flow
Updating}~\cite{JesusBA09}, a fast fault tolerant averaging algorithm
that is, by design, resilient to message loss. 

\bibliography{Practical_Issues}

\end{document}